\documentclass[aps,pra,twocolumn,superscriptaddress]{revtex4-2}
\usepackage{amsmath,amssymb}
\usepackage{bm}

\usepackage{graphicx}
\usepackage[usenames,dvipsnames]{xcolor}
\usepackage{color}
\usepackage[colorlinks=true,linkcolor=blue,urlcolor=blue,citecolor=blue]{hyperref}

\usepackage{orcidlink}

\begin{document}

\title{Static and breathing dynamics of optical bullets in nonlinear GRIN fibers}
\author{Sonali Maity,\orcidlink{0009-0009-2471-5602}}
\affiliation{Department of Physics, Indian Institute of Technology Kharagpur,  Kharagpur  721302, India}
\author{Tiyas Das,\orcidlink{0009-0009-8916-8566}}
\affiliation{Department of Physics, Indian Institute of Technology Kharagpur,  Kharagpur  721302, India}
\author{Anuj P. Lara,\orcidlink{0000-0002-8042-7754}}
\affiliation{Department of Physics, Indian Institute of Technology Kharagpur,  Kharagpur  721302, India}
 \affiliation{Department of Applied Physics and Electronics, Umeå University, Umeå 90187, Sweden} 
\author{Ashis Paul,\orcidlink{0000-0002-9656-9298}}
\affiliation{Department of Physical and Chemical Sciences, University of L'Aquila, Via Vetoio, 67100 L'Aquila, Italy}
\author{ Samudra Roy,\orcidlink{0000-0001-6178-5516}}
\affiliation{Department of Physics, Indian Institute of Technology Kharagpur,  Kharagpur  721302, India}

\date{\today}

\begin{abstract}
We explore the properties of spatiotemporal solitons within nonlinear graded-index (GRIN) optical fibers, specifically addressing their formation, stability, and breathing dynamics. We solve the (3+1)-dimensional nonlinear Schrödinger equation employing a semi-analytical variational approach, which leads to a set of reduced coupled ordinary differential equations describing the bistable bullet dynamics.  The stability of these states is assessed using the \textit{Vakhitov-Kolokolov} criterion and linear stability analysis within the \textit{Kantorovich} optimization framework. The breathing dynamics of the optical bullets are investigated adopting the potential formalism, revealing an oscillatory behavior and self-imaging dynamics in GRIN fibers. The analytical results demonstrate strong agreement with full numerical simulations, validating the robustness of our theoretical model. Our research enhances the understanding of stabilizing self-repeated 3D optical structures, which is beneficial for the development of advanced photonic technologies.
\end{abstract}

\maketitle

\section{Introduction}
Solitons are self-localized nonlinear solitary wave packets that can travel a long distance preserving their shape. The robustness of the soliton structure arises due to a delicate balance between dispersion or diffraction and nonlinearity. In optical media, this balance is achieved by the interplay of group velocity dispersion (GVD) and the Kerr nonlinearity, where the refractive index depends on the light intensity \cite{Hasegawa1973-pl}. This dynamic balance prevents pulse distortion in the time domain and hence enables solitons to retain their structural integrity over exceptionally long distances. These unique features have been the backbone of modern telecommunications and high-speed optical transmission systems, where they support robust, stable, and efficient long-range data transfer~\cite{Hasegawa1973-pl,Mollenauer_1980_9,Hasegawa_22}.

In addition to the conventional temporal solitons that preserve their profile in the temporal domain during propagation, the research interest has recently been focused on the spatiotemporal solitons~\cite{kivshar2003optical} due to their rich physical properties and broad technological relevance \cite{YU1995167,RAGHAVAN2000377,Wise_2015,Cao_2023}. Spatiotemporal solitons, popularly known as \textit{optical bullets}, are the localized wave packets in space and time that are developed due to the intricate balance of diffraction, dispersion, and nonlinearity. However, three-dimensional (3D) bullets undergo critical collapse in a Kerr medium \cite{Silberberg:90}. The formation and stability of these structures are currently at the forefront of nonlinear optics, especially in multimode fiber systems, where the complex mode interactions and higher-dimensional dynamics provide new opportunities for advanced optical control \cite{Desaix1991-yz,Edmundson1992-ws,Cao1994-sg,Hayata1995-po,Malomed1997-vc,Skarka_1997_7,Liu1999-cq}. Owing to their potential applications in ultrafast signal processing \cite{Renninger_2014,Mao_2025}, optical communications, optical digital logic \cite{McLeod1995-ld} and high-power pulse transmission \cite{Malomed_2005}, spatiotemporal solitons continue to be an important area of theoretical and experimental investigation. 

In recent years, significant interest has been devoted to the investigation of spatiotemporal solitons in graded-index (GRIN) fibers due to their unique characteristics of self-imaging \cite{Karlsson:92,Stefano_Longhi_2004,PhysRevAP}, inherent transverse confinement, and enhanced nonlinear wave dynamics \cite{Renninger_2013,Ahsan_18,Krupa_2019_11}. Several theoretical and experimental studies have demonstrated that GRIN fibers offer a unique platform for the formation and guided control of optical bullets due to their inherent transverse confinement and periodic self-imaging properties \cite{Karlsson:92,Stefano_Longhi_2004}. The interplay between spatial confinement, chromatic dispersion, and Kerr nonlinearity in such systems leads to a variety of rich nonlinear dynamical behaviors, including breathing oscillations and stable localized propagation. Understanding these dynamics is therefore essential for the development of advanced multimode photonic devices and ultrafast nonlinear optical technologies.

In this work, we present a structured approach in understanding the dynamics of optical bullet propagation in a nonlinear GRIN fiber.
To investigate the problem, we employ a two-pronged approach consisting of a \textit{Ritz-optimization technique} based on the variational principle \cite{Anderson1999lg} and full 3D numerical simulations of the governing \textit{Nonlinear Schr\"{o}dinger equation}   (NLSE) using the \textit{Split-step Fourier Method} (SSFM)~\cite{Agrawal2012-ym}.
Section.~\ref{sec:theory_model} outlines the theoretical framework and formulates the normalized governing equation by factoring in relevant physical effects. In Section.~\ref{sec:VA}, we apply \textit{variational analysis} (VA) for a semi-analytical solution to the wave propagation problem, deriving coupled ordinary differential equations (ODEs) governing the pulse parameter evolution, obtained through the \textit{Euler–Lagrange}(EL)  formalism. Section.~\ref{sec:Stationary_OB} discusses the existence and properties of stationary optical bullet solutions in the nonlinear GRIN fiber, with a stability analysis with respect to the parametric domain. Section.~\ref{sec:Brathing_OB} establishes the presence of breathing soliton bullets and investigates their pulsating dynamics, using VA results and an effective potential analysis that provides significant physical insight. Additionally, we implement the \textit{Kantorovich approach}~\cite{Vabnitz_2024_11} through Jacobian matrix analysis to delineate the stability regime of these breathing soliton bullets, corroborating all theoretical findings with comprehensive numerical analyses.

\section{Theoretical Model}\label{sec:theory_model}
\begin{figure}
    \centering
    \includegraphics[width=0.95\linewidth]{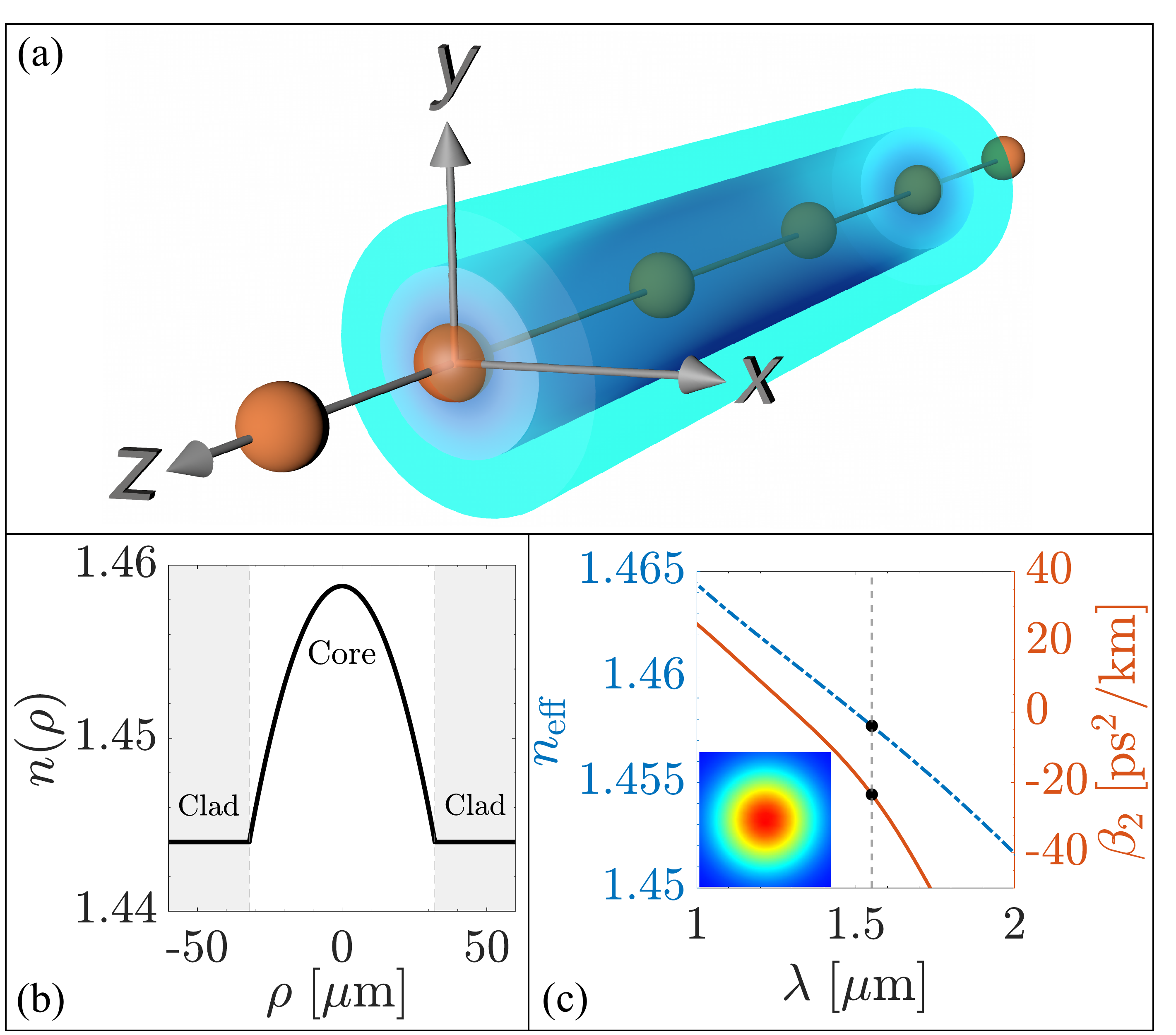}
    \caption{\textbf{(a)} Schematic representation of fundamental bullet generation in a parabolic GRIN fiber. \textbf{(b)} Parabolic radial refractive-index profile inside the core denoted by $n(\rho)$. \textbf{(c)} $n_{\mathrm{eff}}$ as a function of wavelength $\lambda$ (blue dashed dot) and corresponding GVD profile $\beta_2$ (red solid). The vertical grey dashed line denotes the operating wavelength $\lambda_0=1.55\ \mu$m at which $\beta_2=-23.42$ ps$^2$/km. In the inset, we show the fundamental spatial mode distribution inside the proposed GRIN fiber at $\lambda_0=1.55\ \mu$m.}
    \label{fig:3D GRIN}
\end{figure}
We consider a GRIN fiber characterized by a parabolic refractive index profile and incorporating the optical Kerr nonlinearity, with the refractive index expressed as~\cite{maity_2024}
\begin{equation}\label{eq:n_rho}
n(\rho, |E|^2)=n_{\mathrm{core}}\!\left(1-\frac{1}{2}b^{2}\rho^{2}\right)+n_{2}|E|^{2}.
\end{equation}
Here, \(\rho=\sqrt{x^{2}+y^{2}}\) denotes the radial distance from the central axis of the GRIN fiber, and \(n_{\mathrm{core}}\) represents the refractive index at the fiber center (\(\rho=0\)). The parameter \(b\), given by \(b=\sqrt{2\Delta}/a_0\), characterizes the strength of the refractive index gradient, where \(a_0\) denotes the core radius and \(\Delta\) is the relative refractive index difference between the core and the cladding, defined as
\(\Delta=(1-n_{\mathrm{clad}}/n_{\mathrm{core}}),
\)  and \(n_2\) represents the nonlinear Kerr coefficient.
The silica fiber analyzed in this work specifically has a core radius~\cite{Zitelli_2024} $a_0=32\ \mu\mathrm{m}$,  with the value of the Kerr coefficient as~\cite{maity_2024} $n_2=2.7 \times 10^{-20}\ \mathrm{m^2/W}$.
The refractive index is maximum at the core center ($n_{\mathrm{core}}=1.4588$) and parabolically decreases up to the core-cladding interface ($n_{\mathrm{cladding}}=1.4440$) following a parabolic refractive index profile. Fig.~\ref{fig:3D GRIN}(a) illustrates the schematic three-dimensional structure of the GRIN fiber guiding a soliton bullet. Fig.\ref{fig:3D GRIN}(b) and (c)  display the parabolic index profile and GVD profile of the fiber, respectively. The dispersion of the fiber is calculated by utilizing the COMSOL commercial mode-solver software. 
The optical beam inside the fiber is mathematically represented as, $\textbf{E}(\mathbf{r},t) = \hat{\textbf{n}}\text{Re}\left[U (\mathbf{r},t) \exp[i(kz -\omega t)\right]$~\cite{RAGHAVAN2000377}, where $U(\mathbf{r},t)$ is the
slowly varying complex field amplitude, localized both in space and time, and $\hat{\textbf{n}}$ is the unit vector representing the state of the polarization , \ $\omega$ is the carrier frequency, and $k=(\omega/c)n_{\rm core}(\omega)$ is the wavevector of the beam.  Assuming linear polarization and the slowly varying envelope approximation, the spatiotemporal evolution of the optical beam is governed by the (3+1)-dimensional  NLSE derived from the Maxwell's equations~\cite{Ahsan_18}:
	\begin{align}\label{Eq:Governing}
		\frac{\partial U}{\partial z} 
        + \frac{1}{2ik}\nabla^2_{\bot}U
        &+ \frac{i\beta_{2}}{2}\frac{\partial^2 U}{\partial t^2}
        + \frac{i}{2} k b^2 \rho^2 U \nonumber \\
        &- \frac{i \omega n_2}{c}|U|^2U =0.
	\end{align}
Here, $\nabla_{\bot}^2=\partial_x^2 + \partial_y^2$ denotes the transverse Laplacian operator, $t$ is the local time in the co-moving pulse frame, and the parameter \(\beta_2\) corresponds to the GVD. For this work, we neglect the higher-order dispersion.
To simplify the analysis, we normalize Eq. (\ref{Eq:Governing}) by rescaling the parameters as follows:
\begin{equation*}
		\psi = \frac{U}{\sqrt{I_{0i}}}, \quad 
		\rho = r w_{g}, \quad
		\tau = \frac{t}{t_0}, \quad
		\xi = bz,
	\end{equation*}
    where, $I_{0i}$ is the input peak intensity, $w_g$ is the fundamental mode width of a GRIN fiber, defined as $w_g=1/\sqrt{bk}$, and $t_0$ is the temporal width of the input pulse.
    The normalized governing equation then takes the form,
\begin{align}\label{normalised_govn_eqn}
		i\frac{\partial \psi}{\partial \xi} 
		+\frac{1}{2}\left(\frac{\partial^2}{\partial r^2} + \frac{1}{r} \frac{\partial}{\partial r}\right) \psi
		&+ \frac{\delta_{2}}{2} \frac{\partial^2}{\partial \tau^2} \psi 
		-\frac{1}{2 } r^2 \psi  \nonumber \\
		&+\gamma |\psi|^2\psi =0.
	\end{align}
Here, we introduce the dimensionless GVD and nonlinear parameters as
	
	\begin{equation*}
		\delta_{2} = \frac{|\beta_{2}|}{b t_0^2}=\frac{1}{b L_D}, \quad
		\gamma = \frac{\omega n_2 P_{0i}}{A_{\rm eff}bc}.
	\end{equation*} 
 Here $L_D$ is the dispersion length, defined as $L_D = t_0^2/|\beta_2|$, and $P_{0i}$ represents the input power. Employing  COMSOL we calculate the effective refractive index ($n_{\mathrm{eff}}$) of the fundamental mode as a function of wavelength $\lambda$ and further determine the GVD parameter $\beta_2$ as displayed in Fig.\ref{fig:3D GRIN}\textbf{(c)}. At the operating wavelength $\lambda_0=1.55$ $\mu$m, where the fiber's loss is minimum, the GVD parameter is calculated to be $\beta_2 (\lambda_0)=-23.42\ \mathrm{ps^2/km}$. For the given fiber at $\lambda_0=1.55$ $\mu$m, we estimate  
 the effective area  $A_{\mathrm{eff}}=\pi w_g^2=130$ $\mu$m$^2$ and $b=4.5$ mm$^{-1}$.

 \section{Variational Analysis} \label{sec:VA}
This section presents a semi-analytical variational treatment for solving the (3+1)D governing normalized NLSE. Note that one can solve the (3+1)D NLSE numerically, which is time-consuming and requires substantial computational resources. Furthermore, such treatment hinders important physical insights, such as the information on the explicit dependence of system parameters on bullet dynamics.
VA, on the other hand, provides approximate closed-form results carrying significant insight into the problem. 
The success of VA \cite{Anderson_1983_6} relies on the choice of an appropriate \textit{ansatz} function characterizing the properties of the bullet. Additionally, a key assumption within the VA framework is the invariance of the functional form of the chosen \textit{ansatz} during propagation.
To analyze the dynamics of the spatiotemporal optical bullet,  an effective Lagrangian is formulated that corresponds to the (3+1)D NLSE. The application of the EL equations then yields the evolution dynamics of different pulse parameters such as amplitude, width, chirp, phase, etc., in terms of a set of coupled ordinary differential equations (ODEs). The variational framework thereby enables an analytical understanding of the system dynamics through elementary ODE.
Specifically, for this work, we adopt a Gaussian-sech type  \textit{ansatz} where the Gaussian function represents the spatial distribution and a secant-type function captures the temporal profile of the light bullet. Further, for the sake of generality, we consider a parabolic phase-front curvature and a parabolic temporal chirp as a part of \textit{the ansatz} function, which is  mathematically represented as follows:
	\begin{align}
		\psi(r,\tau,\xi)=A_0(\xi) &\operatorname{sech}[\eta(\xi) \tau]
         \exp{\left(-\frac{r^2}{2r_w(\xi)^2}\right)} \nonumber \\
       & \times \exp[i( \phi(\xi) +  d(\xi) r^2 + c(\xi) \tau^2 )],
       \label{ansatz}
	\end{align}
     where $A_0$ denotes the pulse amplitude, $r_w$ represents the radial beam width, and $\eta$ characterizes the inverse of the temporal width of the light bullet. The parameter $d$ corresponds to the spatial phase-front curvature, while $c$ describes the temporal chirp of the pulse. The quantity $\phi$ represents the overall phase of the optical field. All these parameters are considered to be functions of the propagation coordinate $\xi$.
    The Lagrangian density corresponding to Eq.\eqref{normalised_govn_eqn} can be written as follows: 
	\begin{align}
		\mathcal{L}=
		\frac{i}{2} r (\psi \partial_{\xi} \psi^{*} - \psi^{*}\partial_{\xi}\psi)
		+\frac{1}{2} r |\partial_r \psi|^2
		+\frac{\delta_2}{2} r |\partial_{\tau} \psi|^2 \nonumber \\
		+ \frac{r^3}{2}  |\psi|^2
		-\frac{\gamma}{2}r |\psi|^4.
	\end{align}
      Note that the EL equation $\partial_j(\partial_{\psi^*_{j}} \mathcal{L}) - \partial_{\psi^*} \mathcal{L}=0$, with $j= \xi,r,\tau$, applied to this Lagrangian density, reproduces the normalized governing equation Eq. \eqref{normalised_govn_eqn}.
 The reduced Lagrangian is obtained by integrating the Lagrangian density $\mathcal{L}$ over all space as $L = \int_0^\infty 
	\int_{-\infty}^{\infty}\mathcal{L} dr d\tau \;$.
Substituting the \textit{ansatz} function into the Lagrangian density and integrating over the entire space, we obtain the reduced Lagrangian as

\begin{align} 
    L= \frac{A_0^2 r_w^2}{\eta^2} \left( \phi_{\xi} + r_w^2 \eta d_{\xi}+ \frac{\pi^2}{12 \eta}c_{\xi} \right)
   + \frac{A_0^2}{2\eta} \left[ 1+(4d^2+1)r_w^4 \right] \nonumber\\
    -\frac{\delta_{2}A_0^2 r_w^2 \eta}{6}\left(1-\frac{\pi^2 c^2}{\eta^4}\right) 
    - \frac{\gamma A_0^4 r_w^2}{6 \eta}.
\end{align}
Employing EL equation  $\partial_{\xi}(\partial_{X_{\xi}} \mathcal{L}) - \partial_{X} \mathcal{L} =0$, with $X= A_0,\eta,r_w,c,d$ and $\phi$, we  get the evolution equation of the six beam parameters as a set of coupled ODE,
	\begin{subequations}\label{Variation Result}
		
		\begin{align}\label{Es_2}
			\frac{dA_{0}}{d{\xi}} & = - A_0 (2  d  + \delta_{2} c)
			\\
			\label{etas_2}
			\frac{d\eta}{d\xi} & = -2\delta_{2}c \eta
			\\
			\label{rs_2}
			\frac{dr_w}{d\xi} &=2 dr_w 
            \\
            \label{cs_2}
			\frac{dc}{d\xi} &=- 2\delta_{2}  \left ( c^2 - \frac{\eta^4}{\pi^2} \right)    - \gamma\frac{A_0^2\eta^2}{\pi^2}
			\\
			\label{ds2}
			\frac{dd}{d\xi} &= -2 d^2  - \frac{1}{2} \left( 1- \frac{1}{r_w^4}\right) - \frac{\gamma}{6} \frac{A_0^2 }{r_w^2}
			\\
			\frac{d\phi}{d\xi} &=-\frac{1}{r_w^2} - \frac{\delta_{2}}{3}\eta^2 + \frac{7}{12}\gamma A_0^2.
		\end{align}
	\end{subequations}
Note that the phase $\phi$ does not affect the other parameters, reducing the system to 5 independent variables. This set of 5 ODEs is much less time-consuming to solve and provides intricate insight of the problem. Based on these coupled ODEs, in the following sections, we study the stationary and dynamic propagation of the bullets inside a GRIN fiber.

\section{Stationary Optical Bullet Inside Nonlinear GRIN Fiber}
\label{sec:Stationary_OB}
This section examines stationary optical bullet solutions in a nonlinear GRIN fiber, focusing on the conditions necessary for the existence and propagation of stable localized structures without distortion. It also provides a detailed analysis of various characteristics of these stationary states, including their amplitude, spatiotemporal widths, and energy dependence.
    The total energy of the beam is defined as follows:
\(
E=2\pi\int_{0}^{\infty}\int_{-\infty}^{\infty} |\psi|^2 \, r dr \, d\tau .
\)
For the given \textit{ansatz}, as defined in Eq. \eqref{ansatz}, the total energy becomes,
\(
E = \frac{2\pi }{\eta} A_0^2 r_w^2.
\)
The set of Eq.\eqref{Variation Result} yields $\frac{dE}{d\xi}=0$, confirming the energy conservation.
\begin{figure}
    \centering
    \includegraphics[width=\linewidth]{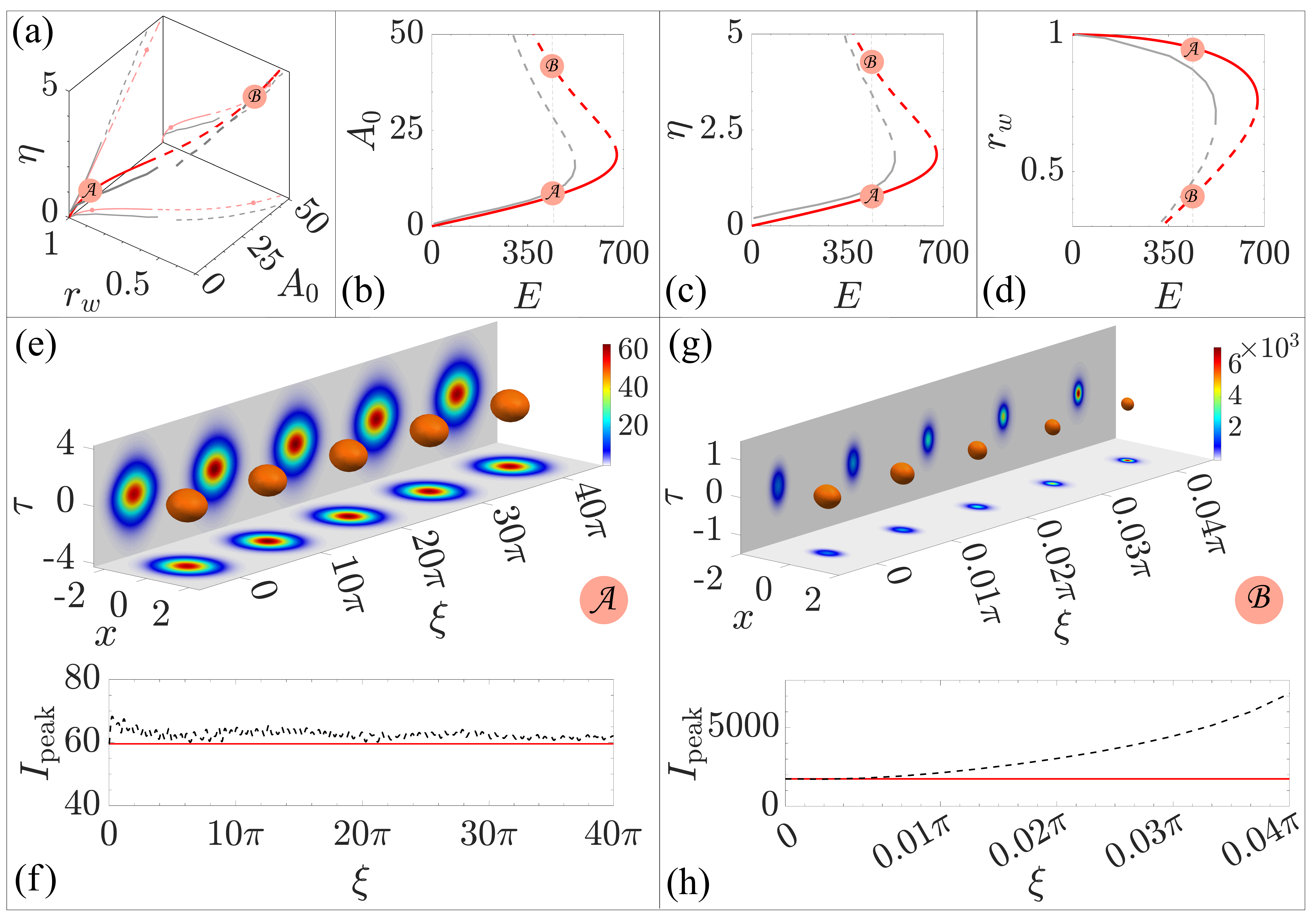}
    \caption{(a) Mutual relation between  $A_0$ , $r_w$ and $\eta$ for static soliton bullets presented in a 3D parametric space. Energy of the stationary bullet as a function of (b) $A_0$, (c) $\eta$ and (d) $r_w$ for $\gamma=0.01$ and $\delta_2=0.5$. The red and black lines correspond to VA and the full numerical results, respectively. The dashed line represents the unstable region of propagation. (e) and (g) depict the dynamics of the light bullet when launched at a point $\mathcal{A}$ and $\mathcal{B}$ on the stability curve, respectively.
    $\mathcal{A}$ and $\mathcal{B}$ corresponds to $A_{0i}=7.72$ and $A_{0i}=41.74$ with identical normalized energy of $E=439$.
     (f) and (h) represent the corresponding intensity variation with propagation distance. In the stability regime, the intensity is almost invariant as predicted by VA (see red line) in contrast to the unstable regime, where exponential growth of amplitude is noticed, which leads to catastrophic collapse of the light beam. 
}
    \label{fig:fig_2}
\end{figure}
The stationary condition can be achieved by making all the $\xi$ derivatives in Eq.\eqref{Variation Result} equal to zero.  Under the chirp-free condition (\textit{i.e} $d=c=0$), the set of  equations yields a relationship between amplitude, spatial width, and inverse temporal width that follows, 
\begin{equation}\label{eq:steady}
	\eta= \sqrt{\frac{3\mathcal{F}}{\delta_2}} ;\quad
  r_w=\sqrt{\mathcal{F}} \left[ -1+\sqrt{1+1/\mathcal{F}^2} \right]^{1/2} \ ,
\end{equation}
where, $\mathcal{F}=A_0^2\gamma/6$.
These relations should be maintained simultaneously during the propagation to achieve a stationary bullet.
Alternatively, the  propagation of the static optical bullet is investigated numerically by assuming a stationary solution of the form $\psi (r,\tau,\xi) = f(r,\tau)e^{iq\xi}$ and substituting it into Eq.~\eqref{normalised_govn_eqn}, we obtain the following equation:
\begin{equation}\label{qequation}
	- q f
	+ \frac{1}{2}
	\left(
	\frac{\partial^2 f}{\partial r^2} + \frac{1}{r}\frac{\partial f}{\partial r}
	\right)
	+\frac{\delta_{2}}{2}\frac{\partial^2f}{\partial \tau^2}
	- \frac{1}{2} r^2 f
	+ \gamma |f|^2 f
	= 0.
\end{equation}  
 Note, Eq. \eqref{qequation} is essentially an eigenvalue problem with an eigenvalue $q$, which we solve numerically, exploiting the \textit{Newton-Raphson} algorithm.
 Fig.~\ref{fig:fig_2} (a) illustrates the relation between the amplitude, spatial width, and inverse temporal width in a 3D parametric space for $\gamma=0.01$ and $\delta_2=0.5$, which attributes a steady-state soliton solution.
The three-dimensional curve, represented by the red solid line, characterizes the parametric dependence obtained using the variational technique. Any point on the solid line should exhibit a stationary bullet solution.
In Fig.~\ref{fig:fig_2} (b)-(d) we demonstrate the dependence of light beam parameters $A_0$, $\eta$, and $r_w$ as a function of stationary bullet energy $E$.
 The numerical results that are obtained by solving Eq. \eqref{qequation} are plotted as the gray lines in Fig.~\ref{fig:fig_2} (a)-(d) which agree well with the analytical predictions (red line).
Most importantly, both analytical (red line) and numerical (gray line) results indicate the existence of bistable stationary states of the optical bullets i.e., two soliton solutions may exist for the same energy. However, it is observed that the stationary solutions are unstable against small perturbations beyond a critical amplitude (dotted line).   
Amplitude-induced self-phase modulation (SPM) is a primary reason for the instability of the stationary bullet at high powers.
 In Fig.~\ref{fig:fig_2}-(e) and (g), we demonstrate the propagation of soliton bullets at the stable (solid lines) and unstable (dotted line) branches indicated by $\mathcal{A}$ and $\mathcal{B}$, on the curves, respectively.  Fig.~\ref{fig:fig_2}-(f) and (h), show the evolution of the peak intensity for stable and unstable bullets, where the numerical results are shown by black dashed lines and the variational results are shown by solid red lines. 
Full numerical analysis, based on the split-step Fourier method \cite{Agrawal2012-ym}, is consistent with the variational prediction obtained by solving the six coupled ODEs in Eq.\eqref{Variation Result} using the fourth-order Runge–Kutta method. 
Both approaches indicate that low-amplitude solitons maintain stability, while high-power solitons demonstrate propagation instability in multiple dimensions, potentially resulting in spatiotemporal wave collapse \cite{Silberberg:90, PARRARIVAS2023171079}. 
This collapse engenders significant compression of spatial and temporal widths, causing a considerable rise in amplitude after a particular propagation distance. A comprehensive stability analysis of these dynamics is addressed in the subsequent section.

\begin{figure}
    \centering
    \includegraphics[width=\linewidth]{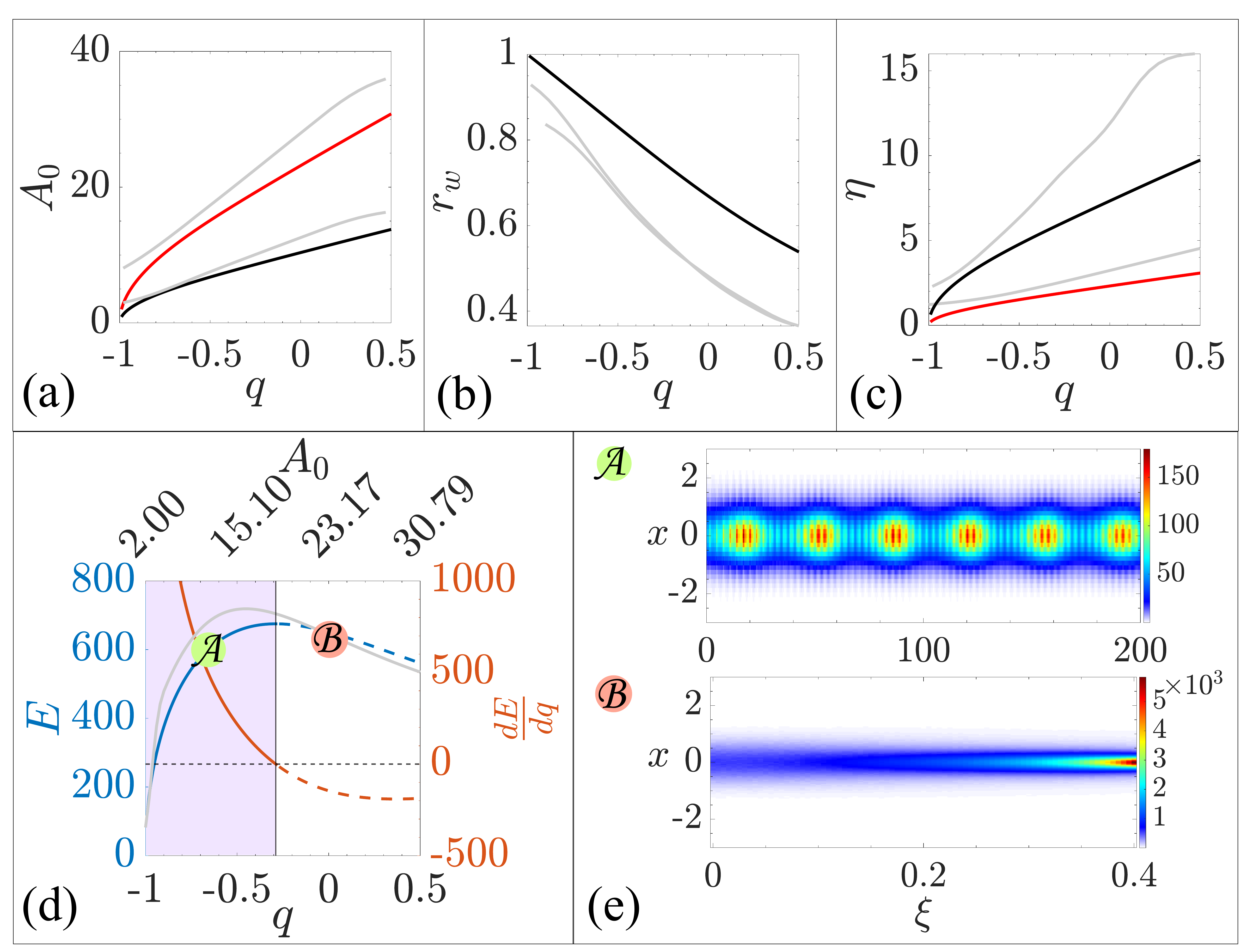}
    \caption{(a)-(c) We plot, $A_0$, $r_w$ and $\eta$ as functions of propagation constant, $q$ where the solid red and black lines correspond to VA results with two sets ($\gamma=0.01$, $\delta_2=0.5$) and  ($\gamma=0.05$, $\delta_2=0.05$), respectively. In all cases, the gray lines indicate numerical results which we obtain by directly solving Eq. \eqref{qequation}.
    (d) Energy $E$ (blue curve, left axis) and its derivative $dE/dq$ (orange curve, right axis) as functions of $q$ for  $\gamma=0.01$ and $\delta_2=0.5$. The upper axis denotes the corresponding values of the amplitude $A_0$. The shaded region represents the stable zone, while the dashed line corresponds to the unstable branch. The horizontal dashed black line indicates $dE/dq=0$ line. In the unstable region, $dE/dq<0$.The two points $\mathcal{A}$ with $A_{0i}=7$ and $\mathcal{B}$ with $A_{0i}=23$ are chosen in the stable and unstable regions, respectively. (e) Projection of the bullet's spatial evolution under an initial perturbation. The bullet propagates steadily when initialized on the stable branch at point $\mathcal{A}$, whereas launching at point  $\mathcal{B}$ leads to catastrophic collapse, demonstrating the instability of the light bullet in the unstable branch.}
    \label{fig:placeholder}
\end{figure}

\subsection{Stability Analysis for Stationary Optical Bullets}
Stability analysis is, in general,  a mathematical diagnostic tool used to determine if the equilibrium of a system is robust enough to withstand external perturbations.
The stability properties of the stationary optical bullet solutions are investigated here by exploiting the \textit{Vakhitov–Kolokolov} (VK) criterion \cite{Vakhitov1973-mt}. Various analytical methods are available in the literature to address the stability in similar systems \cite{PARRARIVAS2023129749,Sun_2024_11}.
 Among these, the VK stability criterion is one of the most widely used methods for assessing soliton stability.  According to the VK condition, the optical bullet is linearly stable if the derivative of the energy $E$ with respect to the propagation constant $q$ remains positive ( $\frac{dE}{dq}>0$) and unstable otherwise. 
  The relationship between $E$ and $q$ can be obtained using the \textit{static Lagrangian} approach, where the Lagrangian density corresponds to Eq. \eqref{qequation} is given as, 
\begin{align}
	\mathcal{L}_{\rm sta}
	=
	\frac{r}{4} \left[\left(2q+r^2\right)f^2 
	+ \left(\frac{\partial f}{\partial r}\right)^2
	+ \delta_{2}\left(\frac{\partial f}{\partial \tau}\right)^2
	-\gamma f^4\right].
\end{align}
Reducing $\mathcal{L}_{\rm sta}$ with a stationary \textit{ansatz} $f(r,\tau) = A_0 \,\text{sech}(\eta \tau) \exp\!\left(-r^2/2r_w^2\right)$, we obtain the  reduced static Lagrangian as follows:

\begin{equation}
	L_{\rm sta} = 
	 \frac{A_0^2 r_w^2}{2\eta} \left(q+\frac{r_w^2}{2} + \frac{1}{2r_w^2}\right) +\frac{A_0^2r_w^2\eta}{12} \left( \delta_2-\frac{\gamma A_0^2}{\eta^2}\right).
\end{equation}
Employing the static EL equation $\frac{\partial L_{\rm sta} }{\partial X}=0$, with $X=A_0, r_w, \eta$, we obtain the beam parameters for the stationary state, 
\begin{subequations}
    \label{12}
    \begin{align} 
        \label{rw^2}
        r_w & = \left[\frac{\sqrt{4q^2 +5}-2q}{5}\right]^{1/2}\\
        \label{A^2}
        A_0 & = \left[\frac{12}{\gamma}\left(q +\frac{\sqrt{4q^2 +5}-2q}{5} \right)\right]^{1/2}\\
        \label{eta^2}
        \eta & = \left[\frac{6}{\delta_2}\left(q + \frac{\sqrt{4q^2 +5}-2q}{5}\right)\right]^{1/2}.
    \end{align}
\end{subequations}
The expressions reveal how system parameters influence the characteristics of a soliton bullet. For example, amplitude $A_0$ only depends on $\gamma$, and the inverse of temporal width $\eta$ only depends on  $\delta_2$. Whereas the spatial beam width $r_w$ is independent of both, $\delta_2$ and $\gamma$. Note that for a realistic solution, the propagation constant has a lower limit, $q=-1$, and no solution can be found for $q<-1$.
The set of Eq. \eqref{12} also yields the relationship between the propagation constant $q$ and the beam energy $E$ as, 
\begin{equation}
	E=E_0 \mathcal{G}(q) \sqrt{\left[ q+\mathcal{G}(q)\right]},
\end{equation}
where, $E_0= \frac{4\pi}{\gamma} \sqrt{6\delta_2}$ and  $\mathcal{G}(q)=-\frac{2}{5}q \left[ 1-\sqrt{1+5/4q^2}\right]$. The instability threshold  of the propagation constant $q_{\rm th}$ is derived by setting $\frac{dE}{dq}=0$, which leads to, 

\begin{equation}
q_{\rm th}=-\frac{1}{2\sqrt{3}} ;~~~~~~~~~A_{\rm th}=\sqrt{\frac{2\sqrt{3}}{\gamma}}. 
\end{equation}
Here $A_{\rm th}$ denotes the instability threshold amplitude. An optical bullet will be unstable if the initial amplitude $A_{0i}$ exceeds $A_{\rm th}$ (i.e $A_{0i}>A_{\rm th}$). Also note that $A_{\rm th} \propto 1/\sqrt{\gamma}$, indicating that GRIN fibers with high nonlinearity do not support stable high-power soliton bullets, which aligns with our numerical findings. In  Fig. \ref{fig:placeholder} (a)-(c) we demonstrate the variation of bullet parameters $A_0$, $r_w$ and $\eta$ as a function of $q$ for two different set of $(\delta_2,\gamma)$. The analytical results (solid red and black lines) are in good agreement with full numerical results (gray lines), which we obtain by solving Eq.\eqref{qequation} numerically. 
In Fig. \ref{fig:placeholder} (d), we demonstrate the $E$-$q$ plot and the stable and unstable soliton bullets (see plot (e)) based on VK criteria. The solitons located in the region $dE/dq>0$ are found to be stable when they are allowed to propagate with random input noise. Solitons with input amplitude beyond the $A_{\rm th}$ are found to be unstable as predicted analytically.

\section{ Breathing Optical bullet inside nonlinear grin fiber}\label{sec:Brathing_OB}
In the preceding sections, we established the existence and investigated the stability of static optical bullets within a nonlinear GRIN fiber, demonstrating that these structures are robust and maintain their initial amplitude and width during propagation.  However, beyond the steady state, a breathing solution may also exist where the parameters of the optical bullet oscillate over propagation distance. The starting point of this analysis should be the normalized governing equation Eq. \eqref{normalised_govn_eqn} which we solve approximately using VA. We employ a $z$-dependent Lagrangian and assume the shape of the optical bullet to be \textit{Gaussian-hyperbolic secant} type, whose parameters are allowed to change over distance. The evolution of the wave parameters are expressed through a set of ordinary coupled differential equations, as demonstrated in Eq. \eqref{Variation Result}. 

\begin{figure}[h!]
    \centering
    \includegraphics[width=\linewidth]{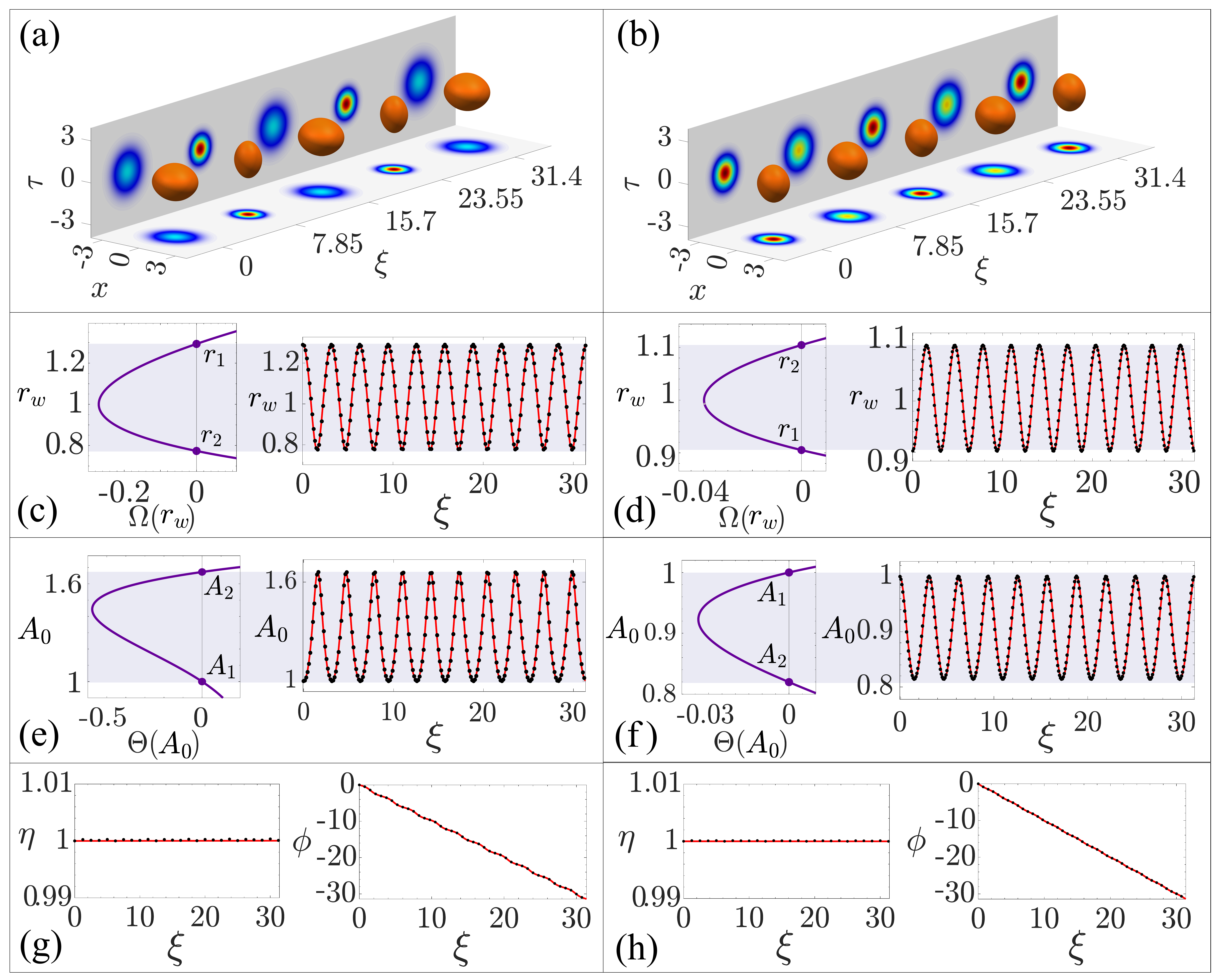}
    \caption{ Dynamics of breathing soliton bullets for (a) initial spectral width $\rho$=$8 \ \mu\mathrm{m}$ ($r_{wi}>1$) and (b) initial spectral width $\rho$=$5.6 \ \mu\mathrm{m}$ ($r_{wi}<1$) with $\gamma=0.01$ and $\delta_2=0.005$. For both the cases $A_{0i}=1$ and $\eta_i=1$. Plot (c) and (d) depict the oscillatory dynamics of spatial beam width $r_w$ guided by the effective potential $\Omega(r_w)$. Similarly plot (e) and (f) illustrate the oscillatory dynamics of amplitude $A_0$ guided by the effective potential $\Theta(A_0)$. The variation of $\eta$ and $\phi$ is shown in plots (g) and (h). In all the cases, the variational results (solid red line) are corroborated with numerical calculations as denoted by black solid dots.}
    \label{fig:potential}
\end{figure}
During breathing, the soliton undergoes periodic compression in the spatial domain while largely preserving its temporal profile.
The oscillatory dynamics of soliton bullet are depicted in two scenarios in Fig. \ref{fig:potential}. In subplot Fig. \ref{fig:potential}(a), the optical bullet is initiated with a spatial width exceeding the fundamental mode width, $w_g$ (\textit{i.e.} $r_{wi}>1$), while in subplot Fig. \ref{fig:potential} (b), it is initiated with a spatial width that is less than $w_g$ (\textit{i.e.} $r_{wi}<1$).  In both cases, the amplitude ($A_0$) and inverse of temporal width ($\eta$) are kept fixed to unity at the input.
Fig.\ref{fig:potential}-(a)-(b) depicts the isosurfaces at intensity $0.65I_{max}$ for the initial spatial width $8 \ \mu\mathrm{m}$ and $5.6 \ \mu\mathrm{m}$, respectively, along with the projection of the intensity at $x$ and $\tau$ plane.

The breathing dynamics of the optical bullet can be further analyzed using the variational results obtained in Eq. \eqref{Variation Result}. Exploiting Eq.\eqref{rs_2} and Eq.\eqref{ds2}, we can obtain a second order differential equation of $r_w$ as
\begin{equation}\label{eq:rw}
	\frac{d^2 r_w}{d \xi^2} +r_w= \frac{\alpha}{r_w^3}.
\end{equation}
where, \(\alpha= \left(1 - \frac{\gamma E  \eta}{6 \pi}\right)\). 
Under suitable input parameters, when the change of temporal width  $\eta$ is minimal over distance, the parameter $\alpha$ can be approximated as a constant, and it is possible to obtain a solution of Eq.\eqref{eq:rw} as follows:
\begin{equation}\label{sol_rw}
	r_{w}(\xi)=\sqrt{\mathcal{C}_{+}-\mathcal{C}_{-}\cos(2\xi)},
\end{equation}
where $\mathcal{C}_{\pm}=\left( \alpha \pm r_{wi}^4\right)/2 r_{wi}^2$ and $r_{wi}$ denotes the initial spatial width. From \eqref{sol_rw} it is easy to show that a steady condition arises when $r_{wi}=\alpha^{1/4}$, which is an alternative representation of Eq. \eqref{eq:steady}. The oscillatory behavior of the breathing soliton is further evaluated employing the effective potential formulation \cite{Skarka_1997_7,Ianetz_2013_4,Tiyas_2026}.
Rearranging Eq. \eqref{eq:rw} and integrating it once with respect to $\xi$, we obtain an equation analogous to that of a particle moving in an effective potential,
\begin{equation}
		\left(\frac{dr_w}{d \xi}\right)^2 + \Omega(r_w) = 0
        \label{rpot}
	\end{equation}
where, the function $\Omega(r_w)$ is defined as,
\begin{equation}
		\Omega(r_w) =  \alpha\left(\frac{1}{r_{w}^2}-\frac{1}{r_{wi}^2}\right) +(r_{w}^2 - r_{wi}^2),
        \label{pot_r}
	\end{equation}
Eq.\eqref{rpot} describes that, for breathing optical bullets, the spatial width $r_w$ oscillates between two turning points (say, $r_1$ and $r_2$) which one can obtain by setting  $\Omega(r_w)=0$. This condition yields spatial width $r_w$ oscillates between two turning points $r_1 = r_{wi}$ and $r_2 = \frac{\sqrt{\alpha}}{r_{wi}}$ around the equilibrium width, $r_{we}=\sqrt[4]{\alpha}$ which is consistent with  Eq. \eqref{sol_rw}. 
Further exploiting the relation, $A_0 = \frac{1}{r_w}\sqrt{\frac{E \eta}{2 \pi}}$ alternatively, we obtain the dynamics of amplitude as, 
\begin{equation}\label{sol_A0}
	A_{0}(\xi)=\frac{\mathcal{H}}{\sqrt{\mathcal{C}_{+}-\mathcal{C}_{-}\cos(2\xi)}},
\end{equation}
where $\mathcal{H}=\sqrt{E\eta/2\pi}$. Eq. \eqref{sol_A0} indicates the beam amplitude $A_0$ oscillates periodically, complementing the spatial beam width $r_w$. The potential approach for $A_0$ leads to another equation,  
\begin{equation}
	\left(\frac{dA_0}{d \xi}\right)^2 + \Theta(A_0) = 0,
\end{equation}
where the function $\Theta(A_0)$ takes the form,
\begin{equation}
	\Theta(A_0) = A_0^4 \left[\frac{\alpha}{\mathcal{H}^4}(A_0^2 - A_{0i}^2) +\left(\frac{1}{A_0^2}-\frac{1}{A_{0i}^2}\right)\right],
\end{equation}
 here $A_{0i}$ denotes the initial amplitude. Likewise, the mathematical condition $\Theta(A)=0$ quantifies the two turning points (say $A_{1}$ and $A_{2}$) within which the amplitude oscillates during propagation.  The turning points $A_{1}$ and $A_{2}$  are calculated to be,  $A_{1} = A_{0i}$ and $A_{2} = \frac{\mathcal{H}^2}{\sqrt{\alpha}A_{0i}}$. Note the equilibrium amplitude is found to be $A_{0e}=\mathcal{H}/\alpha^{1/4}$ for which we have a stationary solution. The potential concept can be further extended to calculate the breathing period of the optical bullet. We consider, $r_w={r}_{we}+\Delta r_w$ where $\Delta r_w$ denotes the deviation from the equilibrium width $r_{we}$. Insert the $r_w$ in Eq. \eqref{rpot} and linearizing it we obtain,
 \begin{equation}\label{eq:second_order_r}
	\frac{d^2 \Delta r_w}{d \xi^2} + w_p^2 \Delta r_w=0,
\end{equation}
where the breathing frequency $w_p$ is defined as, $w_p=\sqrt{\Omega^{''}(r_{we})/2}$. From Eq. \eqref{pot_r} it is straightforward to show that $\Omega^{''} (r_{we})=2+6\alpha/r_{we}^4$ which leads to $w_p=2$ resulting 
 the breathing period $\xi_p=\pi$. Note the breathing period is independent of dispersion and nonlinear parameters, which is consistent with our numerical calculation and previous studies on GRIN fibers \cite{PhysRevA.108.063507,Agrawal:23}.\\
The effective potential, along with the evolution of radial width and amplitude under two input conditions, is illustrated in Fig. \ref{fig:potential}-(c) to (f). The potential plot indicates turning points labeled as $r_1$ and $r_2$  for spatial width \( r_w \), and \( A_1 \) and \( A_2 \) for amplitude \( A_0 \). The oscillatory behavior of the soliton bullet's spatial width and amplitude around their equilibrium positions is essentially governed by Eq. \eqref{sol_rw} and Eq. \eqref{sol_A0} which we derived using VA. 
In plots (g) and (h), we present the numerical and analytical evolution of the inverse temporal width \(\eta\) and phase \(\phi\). The variational predictions, indicated by red solid lines, demonstrate a strong correlation with the numerical results (dotted lines), thereby affirming the validity of the variational approach.

\subsection{Stability Analysis for Breathing Optical Bullets}
\begin{figure}
    \centering
    \includegraphics[width=\linewidth]{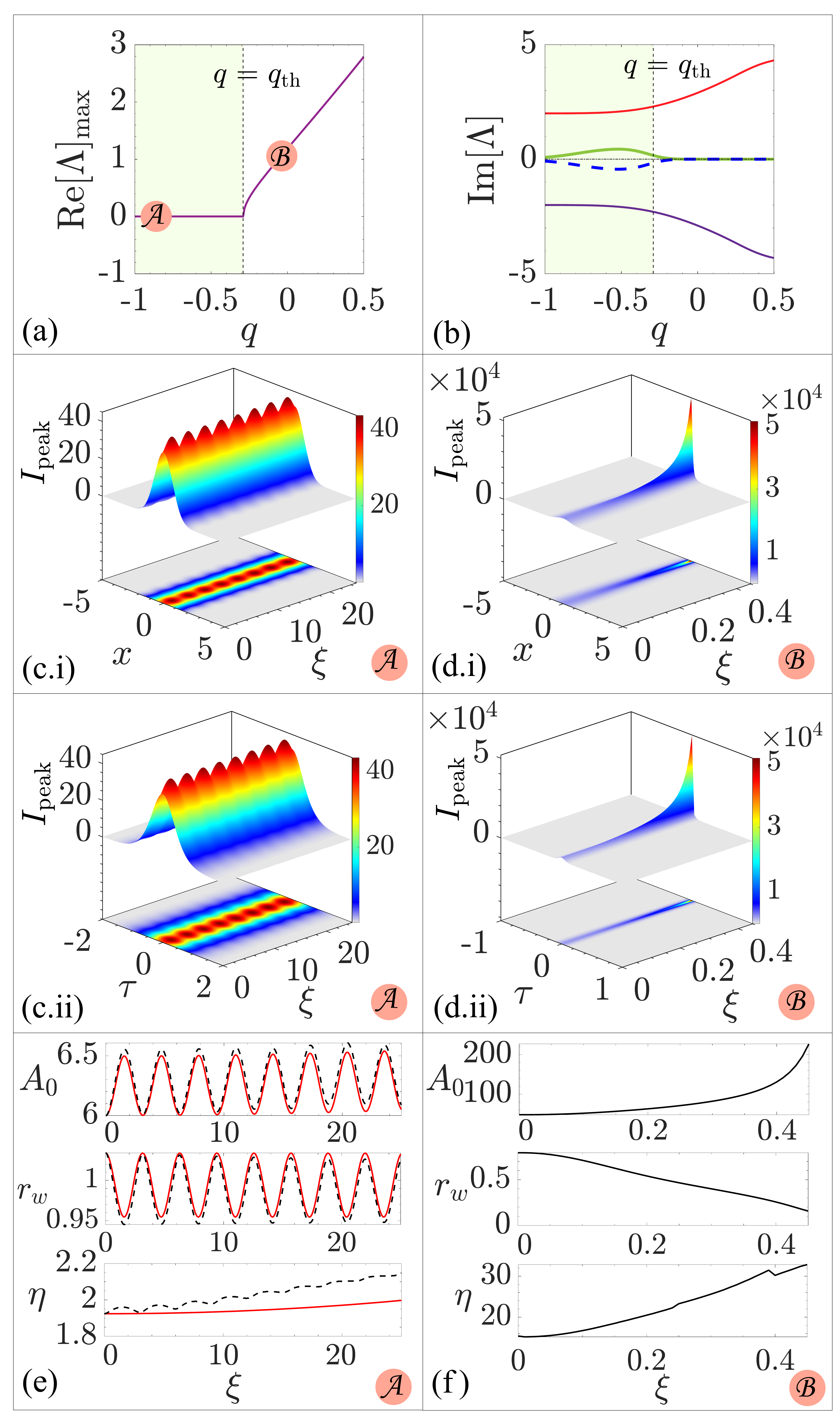}
    \caption{
(a) The real part of the maximum eigenvalue, $\mathrm{Re}[\Lambda]_{\max}$, as a function of $q$ with the system parameters $\gamma=0.0021$ and $\delta_2=0.01$. The points on the instability gain curve marked by $\mathcal{A}$ (with $A_{0i}=6$) and $\mathcal{B}$ (with $A_{0i}=48$) indicate the stable and unstable regimes separated by the threshold point $q_{\rm th}=-0.2886$. Plot (b) shows the imaginary parts of the eigenvalues, $\mathrm{Im}[\Lambda]$, as a function of $q$. Plots (c.i) and (c.ii) display the evolution of a stable breathing structure localized in the $(x,\xi)$ and $(\tau,\xi)$ planes, respectively, which corresponds to point $\mathcal{A}$.
Panels (d.i) and (d.ii) display the catastrophic beam collapses corresponding to point  $\mathcal{B}$ on the instability growth curve. 
Plot (e) illustrates the variation of the amplitude, spatial width, and inverse of temporal width at a point $\mathcal{A}$ when subjected to a small perturbation $\Delta r_w=0.1$. The red line represents the variational result, while the black dashed line corresponds to the numerical result.
Plot (f) illustrates the variation of the amplitude, spatial width, and inverse of temporal width for a point $\mathcal{B}$ under catastrophic beam collapse.} 
    \label{fig:Fig_5}
\end{figure}
A description of breathing soliton bullet propagation can also be developed using the \textit{Kantorovich approach}, in which the \textit{ansatz} parameters are treated as functions of propagation distance \cite{Uzunov_1995, Cerda_1998}. This semi-analytical approach has been successfully employed in previous studies in the context of multidimensional pulse dynamics \cite{YU1995167,RAGHAVAN2000377,Vabnitz_2024_11} .
Let us define the equilibrium points as a column vector $\mu_e=(A_{0e},\eta_e,r_{we},c_e,d_e)^T$, in 5D parameter space, where the suffix $e$ denotes the equilibrium or fixed point.
The perturbative approach allows us to determine how the equilibria of the system react against perturbations of the form $\mu = \mu_e + \epsilon \tilde{\mu}$, where $\epsilon<<1$ and $\tilde{\mu} \equiv (\tilde{A_0},\tilde{\eta},\tilde{r}{_w},\tilde{c},\tilde{d})^{T} $.
Very close to a fixed point $\mu_e$, the dynamics of the system are captured by  linearizing the dynamical system, 
\begin{equation}
    \frac{d\tilde{\mu}}{dz} = \mathcal{J}[\mu_e]\tilde{\mu},
\end{equation}
where $\mathcal{J}[\mu_e]$ is the $5\times 5$ \textit{Jacobian matrix} of the vector field, which is defined by its components as follows:
\begin{equation}
    \mathcal{J}[\mu_e]\equiv \hat{\mathcal{D}}f_{(i,j)}\Big|_{\mu_e} = \left(\frac{\partial f_i}{\partial \mu_j}\right) \Big|_{\mu_e}.
\end{equation}
Here $f_i$ denotes the right-hand side of the set of Eq.\eqref{Variation Result}. For instance, $f_1 =  -A_0 (2  d  + \delta_{2} c)$, $f_2 =  -2\delta_{2}c \eta$  etc.
The $5\times 5$ \textit{Jacobian matrix} is given by,
\begin{equation}
\mathcal{J}=\begin{pmatrix}
0 & 0 & 0 & -\delta_{2}A_{0e} & -2A_{0e} \\
0 & 0 & 0 & -2 \delta_{2}\eta_e & 0 \\
0 & 0 & 0 & 0 & 2r_{we}  \\
-\frac{2 \gamma A_{0e}\eta_e^2}{\pi^2} & \mathcal{J}_{42} & 0 & 0 & 0 \\
-\frac{\gamma A_{0e}}{3 r_{we}^2} & 0 & \mathcal{J}_{53} & 0 & 0
\end{pmatrix}
\end{equation}
where,
$\mathcal{J}_{42} =\frac{2\eta_e}{\pi^2} \left(4\delta_2 \eta_e^2-\gamma A_{0e}^2\right)$, and $\mathcal{J}_{53} = \frac{2}{r_{we}^2} \left(\frac{\gamma A_{0e}^2}{6} -\frac{1}{r_{we}^2}\right)$.
The stability of the breathing soliton bullet under perturbation can be assessed by solving the linear eigenvalue problem $\mathcal{J} {v}=\Lambda {v}$, where $\Lambda$ and ${v}$ are the eigenvalue and eigenvector of $\mathcal{J}$, respectively. The eigen value equation for the $\mathcal{J}$ matrix follows: 
\begin{equation}
\Lambda^5 - \mathcal{P} \Lambda^3 + \mathcal{Q}\Lambda =0,
\end{equation}
where, \(\mathcal{P}=\mathcal{J}_{14} \mathcal{J}_{41} + \mathcal{J}_{15} \mathcal{J}_{51} + \mathcal{J}_{24}\mathcal{J}_{42} + \mathcal{J}_{35}\mathcal{J}_{53}, \) \\
and, 
\(\mathcal{Q}=\mathcal{J}_{14}\mathcal{J}_{41}\mathcal{J}_{35}\mathcal{J}_{53} +
\mathcal{J}_{15}\mathcal{J}_{51}\mathcal{J}_{24}\mathcal{J}_{42}
+ \mathcal{J}_{24}\mathcal{J}_{42}\mathcal{J}_{35}\mathcal{J}_{53}. \)
The corresponding eigenvalues of the instability growth parameter $\Lambda$ are given by,
\begin{equation}
\Lambda = 0,\pm  \left[\frac{\mathcal{P}\pm \sqrt{\mathcal{P}^2-4\mathcal{Q}}}{2}\right]^{1/2}.
\end{equation}
The complete eigen spectrum is illustrated in Fig.~\ref{fig:Fig_5}. Plot (a) and (b) depict the maximum real part (max(Re[$\Lambda$]) and all imaginary parts (Im[$\Lambda$]) of the eigenvalues $\Lambda$, respectively. To examine the dynamical response of the optical bullet, a small perturbation is introduced in the initial spatial width prior to propagation. The point marked by $\mathcal{A}$ lies in the region  $q<q_{th}$ where the real part of $\Lambda$ vanishes, indicating a stable propagation regime. 
An optical bullet launched from this point with an initial normalized amplitude of $A_{0i}=6$, corresponding to an input power of $10~\mathrm{kW}$ and an initial temporal width of $t_0=23~\mathrm{fs}$, undergoes stable oscillatory evolution during propagation. The resulting oscillations, together with their projections along the spatial and temporal directions, are illustrated in plots (c.i) and (c.ii), respectively. In this low-amplitude regime, the numerical results represented by the black dashed lines(Fig.~\ref{fig:Fig_5}-(e)) are found to be in good agreement with the variational predictions shown by the red solid lines, confirming the validity of the VA.
In contrast, point $\mathcal{B}$ is chosen from the region $q>q_{th}$ where the real part of $\Lambda$ becomes non-zero, corresponding to an unstable propagation regime. For this case, the optical bullet is launched with an initial normalized amplitude of $A_{0i}=48$, corresponding to  $10~\mathrm{kW}$ power, keeping the initial width fixed at the previous value. Plot (d.i) and (d.ii) illustrate the dominating SPM-induced catastrophic collapse, which is not captured by the variational treatment owing to significant distortion of the pulse shape.  Therefore, only the numerical results, represented by the black solid lines, are shown for this case (see Fig.~\ref{fig:Fig_5}-(f)).\\

\section{Conclusion}

Our study explores the static and breathing dynamics of spatiotemporal solitons inside a nonlinear GRIN optical fiber, adopting a two-fold approach. The (3+1)-dimensional nonlinear Schrödinger equation that governs the bullet dynamics is solved semi-analytically using VA and further solved numerically using the SSFM algorithm. Exploiting the rigorous VA, which is based on Ritz's optimization principle, we derive 
a set of coupled ODEs describing the dynamic behavior of a soliton bullet 
and elucidate conditions for stationary optical bullet solutions under accessible experimental conditions.  This analytical treatment also predicts the existence of bistable solutions. Our research establishes specific relationships between soliton parameters necessary for equilibrium propagation. To gain a deeper understanding of stability analysis, we perturb the static solution away from the steady state  
and demonstrate the existence of breathing solitons when parameters deviate from equilibrium.   
Stability analysis, employing the Vakhitov–Kolokolov criterion and a Kantorovich-based approach, revealed that beyond a threshold value of amplitude, the bullet becomes unstable.
To get an intuitive idea about the breathing dynamics of the soliton bullets, we further employ effective potential analysis, which sheds more light on the self-imaging oscillations of optical bullets within the GRIN fiber analogous to a point particle trapped in a potential. It is observed that, for the stable regime, the perturbed amplitude and beam width oscillate around equilibrium within a limit defined by the waveguide nonlinearity. Our variational treatment also allows us to determine the oscillation frequency. The analytical predictions are found to be consistent with numerical results, which affirms the reliability of the variational method, suggesting that nonlinear GRIN fibers are promising for stable multidimensional pulse propagation and ultrafast photonic applications, including classical and quantum information technology.

\begin{acknowledgments}
S.M. and T.D. acknowledge the Ministry of Education and Human Resources Development, Government of India, and the Indian Institute of Technology, Kharagpur, for financial support.
\end{acknowledgments}

\bibliography{ref}

\end{document}